\documentclass[12pt,a4paper]{article}
\usepackage{amssymb,amsmath,amscd,epsfig}
\usepackage{graphicx}
\usepackage{latexsym}
\title{ Resonant behaviour in double charge exchange
reaction of $\pi^+$-mesons on the nuclear photoemulsion }

\author{
Yu.A. Batusov\thanks{e-mail:~~yuabat@nusun.jinr.ru}
\hspace*{1mm}$^{\rm 1}$,
T.D. Blokhintseva\thanks{e-mail:~~blokhin@nusun.jinr.ru}
\hspace*{1mm}$^{\rm 1}$,
F. Balestra\thanks{e-mail:~~balestra@to.infn.it }
\hspace*{1mm}$^{\rm 2}$,
M.P. Bussa\thanks{e-mail:~~bussa@to.infn.it}
\hspace*{1mm}$^{\rm 2}$,\\
G. Piragino\thanks{e-mail:~~piragino@to.infn.it }
\hspace*{1mm}$^{\rm 2,3}$,
G.B. Pontecorvo\thanks{e-mail:~~gilpont@nusun.jinr.ru}
\hspace*{1mm}$^{\rm 1,2,3}$,
M.G. Schepkin\thanks{e-mail:~~schepkin@itep.ru}
\hspace*{2mm}$^{\rm 4}$  \\
${\rm ^1}$ {\small\it Joint Institute for Nuclear Research} \\
{\small \it Dubna, Moscow region, 141980, Russia} \\
${\rm ^2}$ {\small\it Dipartimento di Fisica Generale
``A. Avogadro'', Universit\`a di Torino} \\
{\small \it and INFN, Sezione di Torino}  \\
{\small\it  I-10125, Torino, Italy}  \\
${\rm ^3}$ {\small\it Centro Studi e Ricerche "Enrico Fermi",
Roma, Italy}  \\
${\rm ^4}${\small \it A.I. Alikhanov
Institute of Theoretical and Experimental Physics }\\
{\small \it Moscow, 117218, Russia}}

\date{}

\begin{document}

\maketitle

\begin{abstract}
The invariant mass  spectra of the $pp\pi^-$ and $pp$ systems produced in 
the double charge exchange  (DCX) of positively charged pions on photoemulsion 
are analysed. A pronounced  peak is observed in the $pp\pi^-$ invariant mass 
spectrum, while the $M_{pp}$ spectrum exhibits a strong Migdal--Watson effect 
of the proton-proton final state interaction. These findings are in favor of 
the $NN$-decoupled $NN\pi$ pseudoscalar resonance with $T=0$ called $d'$.
\end{abstract}

\section{Introduction}

In this paper the spectrum is analysed of invariant masses of the  
$pp\pi^-$ system, produced in Double Charge eXchange (DCX) reactions of  
$\pi^+$-mesons on the nuclei $^{107}Ag$ and $^{109}Ag$ in photoemulsion; a peak 
has been revealed at $M_{pp\pi^-} \approx 2.05$ GeV. We consider this result 
to confirm the existence of the $d'$, that was proposed in order to explain
the maxima exhibited by the forward angle DCX cross section
$\pi^+~(A,Z) \to (A, Z+2) \pi^-$ at $T_{\pi^+} \sim 50$ MeV 
\cite{BCS}-\cite{Mart}.
The energy of the incident $\pi^+$-mesons in the photoemulsion experiment was 
quite low: $T_{\pi^+} = 40 \div 140 $ MeV, while the resolution in the invariant 
mass $M_{pp\pi^-}$ amounted to about 10 MeV. Thus, the experimental results may 
serve as a critical test of the $d'$-resonance hypothesis. It will be shown that 
an explanation of the results obtained, based on traditional approaches, does 
not provide for reproduction of the maximum in the $M_{pp\pi^-}$ spectrum, 
observed to occupy precisely the region, occupied by the mass of the resonance 
that explained the sharp peaks in the DCX reactions  $\pi^+~(A,Z) \to (A, Z+2) 
\pi^-$  on the isotopes $^7 Li$, $^{12,14}C$, $^{16,18}O$, $^{40,42,44,48}Ca$, 
$^{56}Fe$ and $^{93}Nb$  \cite{BCS}-\cite{Mart}.

In Part 2 of the paper a description is presented of the actual photoemulsion 
experiment. Part 3 presents the results of previous attempts at searching for
the $d'$ together with arguments in favour and against the hypothesis. Then,
Part 4 contains a discussion of possible explanations, based on non-resonant
mechanisms, of the effects observed. Part 5 provides a comparison of calculations 
accounting for the resonance with the experimental results. Part 6 contains the 
conclusions.   

\section{Experimental procedure}

The DCX of $\pi^{\pm}$-mesons was first observed in studies of pion interactions 
with nuclei of photoemulsion \cite{B1}--\cite{B6}. These experiments made use of 
photemulsion chambers, composed of NIKFI--BR layers 400 and 600\,$\mu m$ thick, 
that were exposed to $\pi^+$ and $\pi^-$ beams of the phasotron of the JINR 
Dzhelepov Laboratory of Nuclear Problems;
the momentum spread of the pions in the beams  was $\sim 3\%$. The dimensions of 
these  photoemulsion chambers were chosen so as to provide for 50--140 MeV 
$\pi^-$-mesons being completely stopped in the emulsion.  

The "area" method was applied in searching for stopped $\pi$-mesons in the 
layers of developed emulsion: $\pi^+$-mesons were revealed by the characteristic 
image of the  $\pi^+ \to \mu^+ + \nu_\mu$ decay, while $\pi^-$-mesons were 
recognized by the "$\sigma$-stars". DCX events in chambers exposed to 
$\pi^+$-mesons were identified by stops of $\pi^-$-mesons, and in chambers 
exposed to $\pi^-$-mesons such events were identified by the decays of 
$\pi^+$-mesons.  
The tracks of the stopped mesons were, then, traced back in the emulsion chamber 
up to their departure from stars formed in the emulsion. Only such nuclear 
breakup events were chosen for further analysis, in which: \\
1) the incident pion track was present, \\
2) the angle
between the incident pion track and the beam direction did not exceed $\pm 
3^0$,\\3) the
density of grains along the incident pion track did not differ (within 10$\%$) 
from the average density of grains along tracks of the beam particles. Such an 
analysis of seven photoemulsion chambers (with account of their geometrical 
registration efficiency)  permitted to single out 793 events of the DCX reaction 
\begin{equation}
\label{piplus}
~~\pi^+~+~A ~\to ~\pi^-~+~A'
\end{equation}
and 493 events of the reaction {\footnote { These events will not be further 
discussed, since they cannot be used for the purposes of this paper.}} 
\begin{equation}
\label{piminus}
\pi^-~+~A ~\to ~\pi^+~+~A'~.
\end{equation}

Further analysis of the registered events of $\pi$-meson DCX on the nuclei of 
photoemulsion \cite{B1}-\cite{B6} at initial energies between $40$ and $140$ MeV 
permitted to obtain the following characteristics of the processes studied.   

1. The cross section of $\pi^+$-meson DCX exceeds the cross section of 
$\pi^-$-meson DCX within the entire range of energies studied. If one considers 
DCX reactions to be the result of pions interacting with separate nucleons or 
clusters of nucleons in the nucleus, then the said difference may be attributed 
to several reasons: a) the excess of neutrons in the heavy nuclei ($Ag, Br$) 
present in photoemulsion results in the probability of DCX being higher for 
$\pi^+$-mesons, than for $\pi^-$-mesons. Moreover, additional investigation of 
the dependence of the DCX cross section for both $\pi^+$- and $\pi^-$-mesons 
upon the mass of the target nucleus \cite{B3} reveals that it increases with the 
atomic number of the nucleus; consequently, in photoemulsion this process will 
preferentially proceed on heavy nuclei; b) the emission of secondary 
$\pi^+$-mesons is hindered by the Coulomb barrier of the nucleus. This imposes 
additional restrictions on the DCX of $\pi^-$-mesons.   

2. If all possible reactions are considered of positive $\pi$-meson DCX on 
nuclei present in photoemulsion, then it can be verified that stars with one or 
no prongs cannot be produced on the light nuclei ($C,~N,~O$), and, consequently, 
they result from the interaction of $\pi^+$-mesons with $Ag$ or $Br$ nuclei. The 
same conclusion can be made concerning two-prong events without an electron at the center of the star
\cite{B4}. Consider, for example, the reaction on the C nucleus:
\begin{equation}
\begin{array}{cccc}
\pi^+ + ^{12}C & \to {\pi^-} + {^{12}O} & & \\
 & & \hspace{-1.5cm}\searrow &  \\
 & & p+^{11}N & \\
 & & & \hspace{-1.5cm} \searrow \\
& & & p+^{10}C \\
\end{array}
\end{equation}
with the subsequent beta decay
$^{10}C \to {e^+} + \nu_e + ^{10}B$.

Such a process in photoemulsion looks either like a three-prong star, or like a
two-prong star (when the energy of the residual nucleus $^{10}B$ is low, and it 
leaves no track in the photoemulsion) with a fast electron track departing from 
the center of the star.

Similar reactions take place on N and O nuclei.

3. In the case of $\pi^+$-meson DCX two neutrons of the target nucleus are 
transformed into two protons, that can leave the nucleus. This fact is confirmed 
by the multiplicity distribution of prongs in a star in the case of the 
registered events of $\pi^+$-meson DCX in photoemulsion. In this case, the 
average number of charged particles (without account of the secondary 
$\pi^-$-meson and electrons) per "star" for energies ranging from $40$ to $140$ 
MeV amounts to $2.07 \pm 0.09$.
With account of the restrictions obtained it was possible to identity two-prong
$\pi^+$-meson events of DCX on heavy nuclei of the photoemulsion, the percentage 
of which amounts to $\sim 20 \%$. This result was somewhat unexpected, since the 
Coulomb barrier for $Ag$ and $Br$ nuclei is about 13 MeV, which should 
significantly hinder the departure of two slow secondary protons.

One may assume, that DCX of $\pi^+$-mesons proceeds on a pair of correlated 
neutrons. Then, a noticeable contribution to the reaction should be due
to the $NN\pi$-resonance, $d'$.

In order to test this mechanism experimentally, 224 events of $\pi^+$-meson DCX 
were selected, in which only two protons and a $\pi^-$-meson were found in the 
final state, while no visible tracks of the recoil nucleus or of a fast electron 
were present at the center of the star.

For each of these selected events, the angles were measured between all 
secondary charged particles and the incident $\pi$-meson track, and the momenta 
and energies of the particles were determined from their path ranges. The 
results of these measurements permitted to determine the spectrum of $pp\pi^-$ 
invariant masses (see section 4).
Before initiating a detailed discussion of the resonant mechanism of DCX on 
heavy nuclei we shall present a short review of earlier experimental searches for 
the $d'$ and, also, describe the main properties of the $d'$-resonance. For the 
explanation of DCX on Ag and Br, based on a resonance mechanism, the importance 
should also be noted of the existence in the surface layer of the nuclei of a 
di-neutron, which in nuclei with excess neutrons has been confirmed 
experimentally \cite{B6}.

\section{Searches for the $d'$-resonance in strong and electromagnetic 
processes. Properties of the $d'$-resonance}

\begin{figure}[htb]
\centerline{\epsfig{file=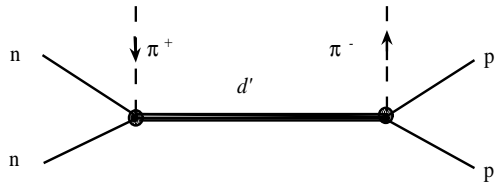, width=7cm, height=3cm}}
\caption{Diagram of DCX reaction} \label{DCX}
\end{figure}

As it was already mentioned, the $d'$-resonance was proposed for explanation of 
the resonance-like behaviour of the $\pi^+$-meson DCX cross section at a forward 
angle for incident pion energies in the vicinity of  $T_{\pi^+} \sim 50$ MeV, 
which turned out to be unexpected from the point of view of conventional nuclear 
models (see, for example, ref. \cite{Naft} and references in ref.\cite{BCS}). 
The reactions are DCX $\pi^+~(A,Z) \to (A, Z+2) \pi^-$, in which the final-state 
nucleus is found either in the ground state (GS) or in the double analog 
isobaric state (DIAS). Thus, it is reasonable to assume (which is further done 
in all papers concerning the DCX) the reaction to take place in the outer shells 
of the nuclei. The idea of DCX as a two-step process resulted in the prediction 
of minima in the region of  $T_{\pi^+} \sim 50$ MeV. This is due to the 
differential single charge exchange cross section in this energy region 
exhibiting deep minima, caused by the destructive interference of $s$- and 
$p$-waves (see review \cite{Heinz}). Contrary to expectations, the first 
experimental studies of DCX revealed quite clear maxima corresponding to a cross 
section exceeding the theoretical predictions \cite{Naft} by two orders of 
magnitude. Subsequent experiments (see \cite{Klaus2} and references therein), 
especially devoted to the investigation of DCX in the region of $T_{\pi^+} = 20 
\div 100$ MeV, confirmed the presence of sharp peaks at the same energy, i.e. at 
$T_{\pi^+} \approx 50$ MeV. Of particular interest is DCX on nuclei with doubly 
closed shells, $^{16}O$ and $^{40}Ca$, for which the DCX cross section, 
according to conventional approaches, should be strongly suppressed as compared 
to DCX on other isotopes (below, the example is presented of the forward angle 
DCX cross section for $^{40}Ca$ from ref. \cite{Klaus1}). But experiments have 
demonstrated that the DCX cross section on all isotopes are of the same order of 
magnitude.  The $NN\pi$-resonance $d'$ with quantum numbers $J^P=0^-,~ T=0$ and 
mass  $M \approx 2m_N + m_\pi + 50$ MeV $\approx 2.06$ GeV is ideal for 
explaining the DCX peaks. Such quantum numbers forbid the decay of $d'$ into a 
pair of nucleons via strong interaction, so the sole decay channel is the decay 
into $NN\pi$ ($pp\pi^-, nn\pi^+$ and $pn\pi^0$). Here, all the three particles 
in the final state are in the $s$-wave with respect to each other.\footnote{The 
electromagnetic decay $d' \to d\gamma$ is also permitted, and it will be dealt 
with below.} Absence of a 2-particle decay via strong interaction signifies that 
the decay width into $NN\pi$ with an energy release $\sim 50$\,MeV should be 
small. Additional suppression of the $d' \to NN\pi$ decay probability is again 
due to the absence of $d'$ coupling to the $NN$-channel: among the diagrams of 
the  $d' \to NN\pi$ decay, there are none, in which the $\pi$-meson is emitted 
from the external hadron line. Hence, from the Adler theorem it follows that the 
3-particle decay amplitude is proportional to the 4-momentum of the pion. In the 
case of an  $s$-wave decay involving a small energy release, this signifies that 
the amplitude contains an additional smallness of the order of $m_{\pi} / m_N$. 
From the above, the diagram describing contribution of the $d'$ to the pionic 
DCX is of the form, presented in fig.\ref{DCX}.

\begin{figure}[htb]
\centerline{\epsfig{file=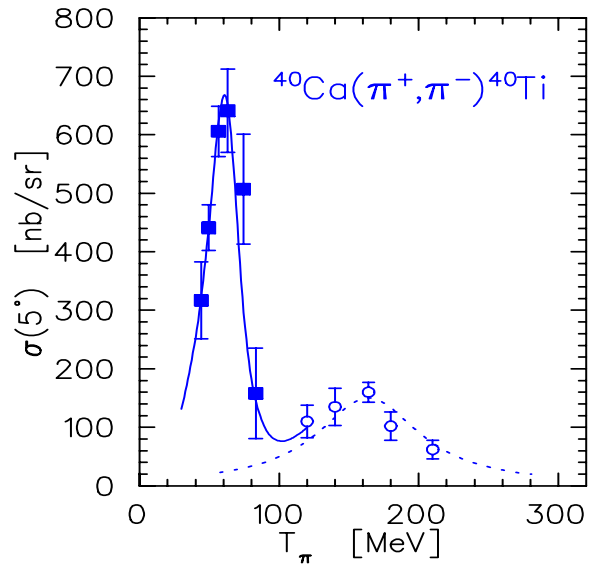, width=8.1cm, height=7.9cm}}
\caption{Energy  dependence of the forward angle DCX cross section on $^{40}Ca$.
The full squares represent experimental measurements from ref.\cite{Klaus1}.
The solid curve shows the $d'$ contribution. The data at $T_\pi \geq 100$ MeV 
(open circles) are from LAMPF \cite{Gilman1}, \cite{Gilman2}. The dotted line 
represents parametrization of the $\Delta\Delta$ process (for more details see
refs.\cite{Klaus1}, \cite{Klaus2}).} \label{Ca}
\end{figure}

For a resonance description of DCX, two other parameters were introduced, 
besides the $d'$ mass: the $d'NN\pi$ coupling constant and the $d'$ decay width 
in a nuclear medium (both the mass and the resonance width in a nuclear medium 
may differ from their vacuum values). As to the $d'$ width in a nuclear medium, 
owing to the reaction $d'N \to 3N$, involving an energy release of 190 MeV, it 
must exceed the $d' \to NN\pi$ decay width in vacuum by an order of magnitude 
\cite{Nef}. The results of fitting the parameters of the resonance over the 
entire set of data on DCX \cite{BCS} are in agreement with calculations based on 
the aforementioned assumptions \cite{Klaus2}, \cite{Nef}. The cross section of 
the reaction $\pi^+$ $^{40}Ca \to ^{40}\hspace{-1mm}Ti~\pi^-$ is presented in 
fig.\ref{Ca} as an example of the DCX cross section behaviour at low energies.

Later, attempts were made \cite{Nus}, \cite{Bill} to explain the peaks 
qualitatively by nuclear medium effects, but no explanation based on 
conventional nuclear physics exists, however, for the whole set of data on DCX, 
and, in particular, for nuclei with $N=Z$.  
A series of experiments in search of the $d'$ has been performed at
JINR \cite{Falomkin}, \cite{Albu} and at TRIUMF \cite{4hetot}, \cite{4hesemi},
where the DCX ${\pi^+} + ^4\hspace{-1mm}He \to pppp~\pi^-$ was measured.
Unlike previous
experiments, DCX on $^4He$ results in the production of four protons in the 
continuum. Therefore, in this reaction the $d'$ production is a threshold 
phenomenon at an initial energy $\sim 80$ MeV: 80 MeV = the resonance energy in 
DCX (50 MeV) plus the binding energy of $^4He$. Besides, the influence of 
nuclear medium effects on the $\pi$-mesons is minimal; one must, however, take 
into account inelastic interactions of the $d'$ with nucleons in the final state 
\cite{Nef}. Taking into account such final-state interactions (like in DCX 
transitions to discrete levels) should result in a significant (approximately by 
an order of magnitude) decrease in the output of "charge-exchanged" $\pi$-mesons, 
produced via the resonance. Thus, the contribution of the resonance to the DCX 
turns out to be much reduced: of, say, 100 produced resonances only a few will 
decay "freely", like in vacuum $d' \to NN\pi$, while the remainder will undergo 
inelastic collisions with the spectator nucleons: $d'N \to 3N$. This hinders 
searches of  $d'$ and may explain \cite{Nef}  the absence of
a threshold effect in DCX on $^3He$, which was also studied \cite{3he} 
experimentally at TRIUMF, after $^4$$He$. However, the total cross section for 
$^4$$He$ as a function of the initial energy exhibits a knee-like behaviour, 
which cannot be reproduced within conventional approaches. The said measurements  
were performed using liquid $^4 He$ \cite{4hetot}; the use of a gas target 
\cite{4hesemi} permitted to measure the invariant mass of the  $pp\pi^-$ system. 
The small statistic and low sensitivity to charged particles with energies below 
$5 \div 7$\,MeV, however, reduces the significance of this exclusive 
measurement.

The most elementary reaction in which $d'$ can be produced
via strong interaction is $pp \to d' \pi^+$. Hence, to search for
an $NN$-decoupled resonance in proton-proton collisions one has
to investigate, for example, two-pion production,
$pp \to pp \pi^- \pi^+$, and to look for the $d'$-signal in the
invariant mass spectrum $M_{pp \pi^-}$.

The first exclusive measurements of two-pion production
in $NN$-collisions \cite{Brod1} at $T_p = 750$ MeV revealed
a narrow peak at $M_{pp \pi^-} \approx 2.06$ GeV.
The authors of ref.\cite{Vorob} also reported evidence of
a resonance-like structure in the $pp\pi^-$ invariant mass
spectrum. However, subsequent experiments, with higher
statistics did not corroborate these findings \cite{Brod2}.

The only reaction where $d'$ can be produced in the $s$-channel
is $\gamma d \to d' \to NN\pi$. In the range of
$2.02 < M_{pp \pi^-} < 2.1$ GeV no narrow structures have been found
with upper limits within a few microbarns \cite{Siodl}.
However, this limit is still an order of magnitude above
the expectation for $d'$ production.

An intriguing possibility to look for the $d'$ would be
pion electroproduction off the deuteron at a large momentum
transfer. At present, however, there are no data which could be
conclusive for the $NN\pi$-resonance with a mass of
about 50 MeV above the threshold.

From all available experimental results of $d'$ searches
one can conclude that no unambiguous answer to the question
concerning the $d'$ existence has been found yet.

\section{Analysis of the $pp\pi^-$ invariant mass spectrum}

In this section we present the measured distribution
of $M_{pp \pi^-}$, and analyse possible reaction channels
of the pionic double charge exchange on nuclear photoemulsion
accompanied by the emission of two protons.

As it was already explained in section 2, the reaction
mainly takes place
on heavy isotopes ($Ag$ and $Br$). Thus, for example, in the
case of $Ag$ isotopes, the following reactions can contribute
to the double charge exchange, in which final states contain
only stable isotopes:
\begin{equation}
\pi^+ + ^{107} \hspace{-2mm} Ag
\to ^{105} \hspace{-2mm} Ag + pp\pi^-
\label{107-1}
\end{equation}
\begin{equation}
\pi^+ + ^{109}\hspace{-2mm}Ag \to ^{107}\hspace{-2mm}Ag + pp\pi^-
\label{109-1}
\end{equation}
\begin{equation}
\pi^+ + ^{109}\hspace{-2mm}Ag \to ^{105}\hspace{-2mm}Ag + pp\pi^- +2n
\label{109-2}
\end{equation}
\begin{equation}
\pi^+ + ^{107}\hspace{-2mm}Ag \to ^{103}\hspace{-2mm}Ag + pp\pi^- +2n
\label{107-2}
\end{equation}
\begin{equation}
\pi^+ + ^{107}\hspace{-2mm}Ag \to ^{54}\hspace{-1mm}Cr +
^{51}\hspace{-1mm}V +pp\pi^-
\label{107-3}
\end{equation}

Similar reactions can take place on $Br$ isotopes:
\begin{equation}
\pi^+ + ^{81}\hspace{-1mm}Br \to ^{79}\hspace{-1mm}Br
+ pp\pi^-
\label{81-1}
\end{equation}
\begin{equation}
\pi^+ + ^{81}\hspace{-1mm}Br \to ^{77}\hspace{-2mm}Br +
pp\pi^- + 2n
\label{81-2}
\end{equation}
\begin{equation}
\pi^+ + ^{79}\hspace{-1mm}Br \to ^{77}\hspace{-2mm}Br + pp\pi^-
\label{79-1}
\end{equation}
\begin{equation}
\pi^+ + ^{79}\hspace{-1mm}Br \to ^{37}\hspace{-1mm}Cl +
^{40}\hspace{-2mm}Ar +pp\pi^-
\label{79-2}
\end{equation}

All other reactions on $Ag$ and $Br$ isotopes induced
by $\pi^+$ (with the $pp\pi^-$ subsystem in the exit channel)
produce radioactive isotopes, and are discarded.

\begin{figure}[htb]
\centerline{\epsfig{file=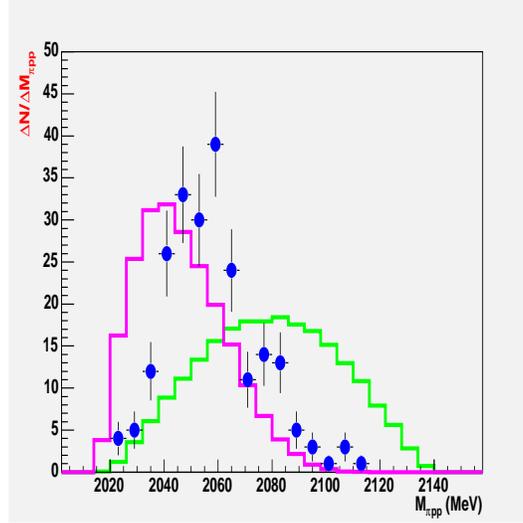,
width=7cm, height=7cm}}
\caption {$pp\pi^-$ invariant mass distribution. The measured
spectrum is shown by dots with error bars. Right and left
histograms are results of MC simulations for DCX transitions
$\pi^+ + ^{107} \hspace{-2mm} Ag
\to ^{105} \hspace{-2mm} Ag + pp\pi^-$ and
 $\pi^+ + ^{107}\hspace{-2mm}Ag \to ^{103}\hspace{-2mm}Ag +
 pp\pi^- +2n$,  respectively. }
\label{pppi-exp}
\end{figure}

Shown in fig.\ref{pppi-exp} are experimental data in comparison
with $M_{pp\pi^-}$ distributions for reactions (\ref{107-1})
and  (\ref{107-2}) (right and left histograms, respectively).

The ${pp\pi^-}$ invariant mass distributions for the reactions
(\ref{109-1}), (\ref{81-1}) and (\ref{79-1}) do practically
not differ from the distribution for the reaction (\ref{107-1}).
The histograms for reactions with extra two neutrons
(\ref{109-2}), (\ref{107-2}) and (\ref{81-2}) are also very
close to each other. The distribution over $M_{pp\pi^-}$ for the
cases when production of the ${pp\pi^-}$ subsystem is
accompanied by a breakup of the rest of the nucleus
(reactions (\ref{107-3}) and (\ref{79-2})) appears to be very
similar to the $M_{pp\pi^-}$-spectrum for reactions
(\ref{107-1}), (\ref{109-1}), (\ref{81-1}) and (\ref{79-1}),
which is not surprising, because the energy carried away
by heavy fragments is negligible.

That is the reason to present in fig.\ref{pppi-exp} two
kinds of MC simulations for those reactions, in which the
$M_{pp\pi^-}$ spectra are essentially different.
The histograms shown in fig.\ref{pppi-exp} represent phase space
distributions for the reactions listed above.
As it can be seen, none of them (neither their combination) can
satisfactorily reproduce the observed $M_{pp\pi^-}$ spectrum.
The experimental resolution for the $M_{pp\pi-}$ invariant mass 
distribution is $\sim 10$ MeV; it has been taken into account 
in our simulations shown in  fig.\ref{pppi-exp}.

Taking into account $NN$  FSI (see section 5)
does not allow to achieve agreement between the simulated
results and experimental data. Inclusion of the $pp$ FSI
shifts the left histogram to the region of even smaller values
of $M_{pp\pi^-}$, while $nn$ FSI practically leaves the shape
of the left histogram and the position of its maximum practically
intact. As for the right histogram, it stays as wide as in
fig.\ref{pppi-exp} after taking into consideration $NN$ FSI.

Another possibility could be production of the ${pp\pi^-}$
system via the $d'$-resonance,

\begin{equation}
\pi^+ + ^{(A)}\hspace{-2mm}Ag \to ^{(A-2)}\hspace{-2mm}Ag + d'
\label{photo-res}
\end{equation}
with subsequent decay of $d'$ into $pp \pi^-$. Similar
production of the $d'$ can, of course, take place on $Br$
isotopes.

The resonance mechanism should also contribute to production
of the $pp\pi^-$ system when produced together with two extra
neutrons:
\begin{equation}
\pi^+ + ^{109}\hspace{-2mm}Ag
\to ^{105}\hspace{-2mm}Ag + d' +2n
\label{109-R}
\end{equation}
which is the resonance mechanism of the reaction
(\ref{109-2}) and (\ref{107-2}). 
If $d'$ really exists, its effect in the
$pp\pi^-$ invariant mass spectrum should be similar
for reactions (\ref{photo-res}) and (\ref{109-R}).
Threshold energies for production of the $d'$ in reactions
(\ref{photo-res}) for $A=109$ and 107,  and in reaction
(\ref{109-R}) are different and are readily calculated.
This was automatically taken into consideration when the 
$M_{pp\pi^-}$ distributions were averaged over the spectrum of 
the incident $\pi^+$-mesons, which is nearly flat.
 This averaging takes into account that for any particular value of
$M_{pp \pi^-}$ there is a certain (minimal) value of the energy of the
incident $\pi^+$.

\begin{figure}[htb]
\centerline{\epsfig{file=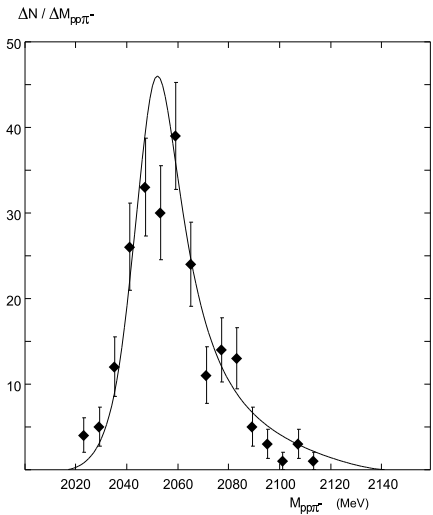, width=7cm, height=8cm}}
\caption{ Resonant description of $pp\pi^-$ invariant mass
spectrum in comparison with experimental data for reaction
${\pi^+}+~^{107}$Ag$\to$ $ ^{105}$Ag$~~+ pp~\pi^-$.}
\label{Res}
\end{figure}

A similar procedure for averaging  the $M_{pp\pi^-}$ spectrum
was also performed for non-resonant mechanisms discussed
above (fig.\ref{pppi-exp}). The resonance contribution is shown in
fig.\ref{Res}.

As it is seen, the resonance contribution provides a reasonable fit
of the experimentally observed ${pp\pi^-}$ invariant mass
distribution. The mass of the
resonance required for description of the experimental
$M_{pp\pi^-}$ spectrum equals $\approx$ 2.05 GeV. This value
practically coincides with the one deduced from pionic DCX
described in section 3. The
width of the resonance, fitted from the experimental data shown
in fig.\ref{Res}, is about 20 MeV. It deserves mentioning
that the vacuum width of the $d'$ deduced from the data on DCX
transitions to the discrete levels is only $1/2 \div 1$ MeV,
while its total width in a nuclear medium was found to be
$15 \div 20$ MeV. In the reaction under consideration there
must also be nuclear medium effects, and hence the observed
width of the resonance must be closer to its width in a nuclear
medium than to its vacuum width. Thus, the parameters of
the resonance required for fitting the $M_{pp\pi^-}$ spectrum
coincide with the corresponding parameters deduced from DCX
transitions to discrete levels.
It must be mentioned that the fitted width of the resonance
is to be understood as an effective width, which takes into account 
the vacuum width, the width of the resonance in nuclear medium and 
the broadening of the $M_{pp\pi^-}$ distribution due to the 
experimental resolution. Thus the physical width (which is mainly
governed by nuclear medium effects) is somewhat smaller than
the one used to  calculate the resonant contribution (solid
curve in fig.\ref{Res})

\section{$pp$ final state interaction}

\vspace{1cm}

An effective way to study more about the reaction mechanism of
$pp\pi^-$ production in the double charge exchange reaction is to
investigate the $pp$-invariant mass distribution. As it is well known,
the low mass side of the $M_{pp}$-spectrum must be affected by the
Migdal-Watson effect of $NN$ final state interaction (FSI)
\cite{Watson}, \cite{Migdal} when the two nucleons are in a relative
$s$-wave. The enhancement factor of the $M_{pp}$-spectrum
can be calculated from:

\begin{equation}
F=\left | 1+ A \right|^2 F_C(pa_c),
\label{f}
\end{equation}
$$
A=\frac{\rho^{-1}}
{-a_s^{-1} +\frac{1}{2}r_0 p^2-\frac{2}{a_c}h(pa_c)
-ipF_C(pa_c)}.
$$

Here $a_s = -7.8$ fm is the $pp$ scattering length, $r_0 =2.8$ fm
is the effective range, $p$ is the momentum
of either proton in their c.o.m., $p=\sqrt{M_{pp}^2 / 4 -m_p^2}$;

\begin{equation}
F_C(x)=\frac{2\pi / x}{e^{2\pi / x}-1}
\end{equation}

\begin{equation}
h(x)= \frac{1}{x^2} \sum_{n=1}^\infty
\frac{1}{n(n^2+x^{-2})} -\gamma + \ln (x)~,
\label{h}
\end{equation}
$\gamma \approx 0.577$ is the Euler constant,
and $a_c = 57.5$ fm.

The formula for calculation of the FSI effect takes into account Coulomb
repulsion. The
enhancement factor given by eq.(\ref{f}) contains a sole parameter $\rho$,
the value of which depends on the particular reaction mechanism. The
smaller the interaction region, in which the two nucleons are produced,
the smaller the parameter $\rho$, and the stronger the effect of NN FSI.
Thus, for example, if the two-proton source is considered
pointlike, the parameter $\rho$ can be as small as $\sim 1$ fm.
This estimate can be found from the Yamaguchi model for
the $NN$ wave function in continuum \cite{Yamaguchi}.

\begin{figure}[htb]
\centerline{\epsfig{file=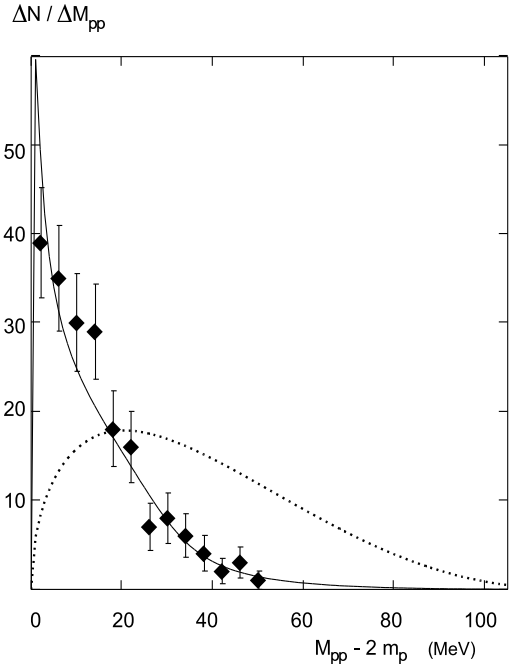, width=9cm, height=11cm}}
\caption{$pp$ invariant mass distribution in reaction
${\pi^+}+~^{107}$Ag$\to$$ ^{105}$Ag$~~+ pp~\pi^-$:
experimental points and fit of data based on
eq.(\ref{f}). The dotted curve is the phase space distribution for
reaction ${\pi^+}+ ^{107}$Ag$\to$$^{105}$Ag$+ pp~\pi^-$.}

\label{pp}
\end{figure}

Shown in fig.\ref{pp} is the $pp$ invariant mass distribution.
Experimental data demonstrate a rather strong enhancement of the
low mass side of the $M_{pp}$ spectrum. This may point to
significant $pp$ FSI if the DCX transition is dominated by
the reaction mechanism without emission of two neutrons.


The dotted
curve is the phase space  distribution of $M_{pp}$ for reaction
(\ref{107-1}). The solid curve takes into consideration $pp$ FSI
calculated using eq.(\ref{f}).
The parameter $\rho$ needed to describe the observed FSI effect
was found to be: $\rho = 1.1$ fm.

Such a strong FSI effect can hardly be expected for
conventional (non-resonant) production of the $pp\pi^-$ system.

At the same time, it is quite reasonable for the resonant
production of the $pp\pi^-$ system, if $d'$ is basically a
6$q$-object with a typical hadronic size $\sim 1 \div 1.5$ fm.

\section{Conclusion}
224 positive pion DCX events on heavy nuclei in
photoemulsion have been analysed and the invariant
{\bf $M_{pp\pi^-}$} spectrum has been shown to indicate
a resonant behaviour at low energies.

\section{Acknowledgments}

We are grateful to I.Gnesi for help.

This work was partially supported  by the
grant SH-2328.2003.2 and by the RFBR grant 05-02-17471.


\begin{thebibliography}{99}

\bibitem{BCS}
R. Bilger, H. Clement, M. Schepkin, Phys. Rev. Lett.,
{\bf 71} (1993) 42.

\bibitem{Betal} R. Bilger, H. Clement, K. Foehl, K. Heitlinger, C. Joram, 
W. Kluge, M. Schepkin, G.J. Wagner, R. Wieser, R. Abela, F. Foroughi, D. Renker , 
Z. Phys., {\bf A343}.(1992) 491.

\bibitem{Mart}
B. Martemyanov, M. Schepkin, JETP Lett., {\bf 53} (1991) 139.

\bibitem{B1}
Yu.A. Batusov, S.A. Bunyatov, V.M. Sidorov, V.A. Yarba, JETP, {\bf 46} (1964) 817 
(in Russian).

\bibitem{B2}
Yu.A. Batusov, S.A. Bunyatov, V.M. Sidorov, V.A. Yarba, Sov.J.Nucl.Phys., 
{\bf 1} (1965) 271.

\bibitem{B3}
Yu.A. Batusov, S.A. Bunyatov, V.M. Sidorov, V.A. Yarba, Sov.J.Nucl.Phys., 
{\bf 3} (1966) 223.

\bibitem{B4}
Yu.A. Batusov, S.A. Bunyatov, V.M. Sidorov, V.A. Yarba,
Yad. Fiz., {\bf 6} (1967) 998 (in Russian).

\bibitem{B5}
Yu.A. Batusov, V.I. Kochkin, V.M. Malzev, Yad. Fiz., {\bf 6} (1967) 158 (in 
Russian).

\bibitem{B6}
R. Kalpakchieva, Yu.E. Penionzhkevich, H.G. Bohlen, Phys.Part.Nucl., {\bf 29} 
(1998)341.

\bibitem{Naft} N. Auerbach, W.R. Gibbs, Joseph N. Ginocchio, W.B. Kaufmann, 
Phys. Rev.,{\bf C 38}  (1988) 1277.

\bibitem{Heinz} H. Clement, Prog. Part. Nucl. Phys.,
{\bf 29}  (1992)  175.

\bibitem{Nef} A.V. Nefediev, M.G. Schepkin, H.A. Clement,
 Phys. Rev., {\bf C 67} (2003) 015201.

\bibitem{Klaus1}
K. Foehl, R. Bilger, H. Clement, J. Gr\"ater, R. Meier, J. P\"atzold, D. 
Schapler, G. J. Wagner, O. Wilhelm, W. Kluge, R. Wieser, M. Schepkin, R. Abela, 
F. Foroughi, D. Renker, Phys. Rev. Lett., {\bf 79} (1997) 3849.

\bibitem{Klaus2} 
J. Draeger, R. Bilger, H. Clement, M. Cr\"oni, H. Denz, J. Gr\"ater, R. Meier, 
J. P\"atzold, D. Schapler, G.J. Wagner, O. Wilhelm, K. F\"ohl, M. Schepkin , 
Phys. Rev., {\bf C 62} (2000) 064615.

\bibitem{Gilman1}R. Gilman, H.T. Fortune, M.B. Johnson, E.R. Siciliano, H. Toki, 
A. Wirzba and B.A. Brown, Phys. Rev., {\bf C 34} (1986) 1895.

\bibitem{Gilman2}R. Gilman, H. T. Fortune, J. D. Zumbro, C. M. Laymon,  
                 G. R. Burleson, J. A. Faucett, W. B. Cottingame,  
                 C. L. Morris 
                 Peter A. Seidl, C. Fred Moore, L. C. Bland, Rex R. Kiziah
                 S. Mordechai,  
                 Kalvir S. Dhuga,  
                 Phys. Rev., {\bf C 35} (1987) 1334.

\bibitem{Nus} M. Nuseirat, M. A. K. Lodhi, M. O. El-Ghossain, W. R. Gibbs,
W. B. Kaufmann , Phys. Rev.,  {\bf C 58} (1998) 2292.

\bibitem{Bill} H.C. Wu, W.R. Gibbs, Phys. Rev., {\bf C 62}
(2000) 044614.

\bibitem{Falomkin} I.V. Falomkin, M.M.Kulyukin, V.I.Lyashenko, G.B.Pontecorvo, 
Yu.A. Shcerbakov, C. Georgescu, A. Mihul, F. Nichitiu, A. Seraru, 
G. Piragino, Nuovo Cimento, {\bf A 22} (1974) 333.

\bibitem{Albu}
 I.V. Falomkin, V.I. Lyashenko, G.B. Pontecorvo, Yu.A. Shcerbakov, M. Albu, 
A. Mihul, F. Nichitiu, A. Seraru, F, Balestra, R. Garfagnini,G. Piragino,
Lett. Nuovo Cim., {\bf 16} (1976) 525.

\bibitem{4hetot}
J. Gr\"ater , R. Bilger, H. Clement, R. Meier, G. J. Wagner, E. Friedman,
M. Schepkin, P. A. Amaudruz, L. Felawka, D. Ottewell, G. R. Smith, A. Ambardar,
 G. J. Hofman, M. Kermani, G. Tagliente, F. Bonutti, P. Camerini, N. Grion, 
R. Rui, P. Hong, E. L. Mathie,  R. Tacik, J. Clark, M. E. Sevior, O. Patarakin,
 Phys. Rev., {\bf C 58} (1998) 1576.

\bibitem{4hesemi}
J.L. Clark,M. E. Sevior,
H. Clement, J. Gr\"ater, R. Meier, G. J. Wagner, P.-A. Amaudruz, L. Felawka, 
G. J. Hofman, D. Ottewell, G. R. Smith, A. Ambardar, M. Kermani, G. Tagliente,
P. Camerini, E. Fragiacomo, N. Grion,  R. Rui, E. L. Mathie, R. Tacik, 
D. M. Yeomans, E. F. Gibson, J. T. Brack, M. Schepkin,  Phys. Rev., 
{\bf C66}  (2002)  054606.

\bibitem{3he} J. Gr\"ater, R. Bilger, H. Clement, R. Meier, J. P\"atzold, 
G.J. Wagner, E. Friedman, E.L. Mathie, R. Tacik, M. Yeomans, P.A. Amaudruz, 
L. Felawka, D. Ottewell, K. Raywood, G.R. Smith, G.J. Hofman, B. Jamieson, 
M. Kermani, G. Tagliente, P. Camerini, E. Fragiacomo, N. Grion, R. Riu, J. Clark, 
G. Moloney, M.E. Sevior, A. Nefediev, M. Schepkin, E.F. Gibson, O. Patarakin,
H. Staudenmaier, S.N. Filippov, Yu.K. Gavrilov, T.L. Karavicheva ,   
Phys. Lett., {\bf B471} (1999) 113.

\bibitem{Brod1} W. Brodowski, R. Bilger, H. Calen, 
H. Clement, C. Ekstroem, K. Foehl, K. Fransson, L. Gustafsson, S. Haeggstroem, 
B. Hoistad, A. Johansson, T. Johansson, K. Kilian, S. Kullander, A. Kupsc, 
G. Kurz, P. Marciniewski, B. Morosov, J. Moehn, A. Moertsell, W. Oelert, 
V. Renken, R. Ruber, M.G. Schepkin, U. Siodlaczek, J. Stepaniak, A. Sukhanov, 
A. Turowiecki, G.J. Wagner, Z. Wilhelmi, J. Zabierowski, A. Zernov, J. Zlomanczuk, 
Z.Physik, {\bf A355} (1996) 5.

\bibitem{Vorob} L.S. Vorobyev, Yu.G. Grishuk, 
Yu.V. Efremenko, M.V. Kossov, S.V. Kuleshov, G.A. Leksin, N.A. Pivnyuk, 
A.V. Smirnitsky, V.B. Fedorov, B.B. Shvartzman, S.M. Shuvalov, M.G. Schepkin, 
Phys.Atom.Nucl., {\bf 61} (1998) 771 and JEPT Lett., {\bf 59} (1994) 77.

\bibitem{Brod2} W. Brodowski, J. P\"atzold, R. Bilger,H. Calen, C. Ekstr\"om, 
K. Fransson, J. Greiff, S. Haggstrom, B. Hoistad, J. Johanson, A. Johansson, 
T. Johansson, K. Kilian, S. Kullander, A. Kupsc, P. Marciniewski, B. Morosov, 
W. Oelert, R.J.M.Y. Ruber, M. Schepkin, W. Scobel, J. Stepaniak, A. Sukhanov, 
A. Turowiecki, G.J. Wagner, Z. Wilhelmi, J. Zabierowski, J. Zlomanczuk,
Phys. Lett., {\bf B550}  (2002)  147.

\bibitem{Siodl}U. Siodlaczek, P. Achenbach, 
J. Ahrens, H.J. Arends, R. Beck, R. Bilger, H. Clement, V. Hejny, J.D. Kellie, 
M. Kotulla, B. Krusche, V. Kuhr, R. Leukel, J.C. McGeorge, V. Metag, R. 
Novotny, V. Olmos de Leon, F. Rambo, M. Schepkin, A. Schmidt, H. Stroher, 
G.J. Wagner, T. Walcher, J. Weiss, F. Wissmann, M. Wolf, Eur. Phys. J., 
{\bf A9} (2000) 309.
\bibitem{Watson}  K.M. Watson.  Phys. Rev., {\bf 88}  (1952) 1163.

\bibitem{Migdal}  A.B. Migdal, Sov.Phys. JETP, {\bf 1} (1955) 2.

\bibitem{Yamaguchi}  Y. Yamaguchi, Phys. Rev., {\bf 95}  (1954) 1628.

\end{thebibliography}
\end{document}